\documentclass[aps,prd,eqsecnum,showpacs]{revtex4}
\usepackage[dvips]{color,graphicx}
\usepackage{amsfonts}
\usepackage{amssymb}

\begin{document}

\title{
Bremsstrahlung Effects around Evaporating Black Holes}

\author{
$^{1}$Don N. Page\footnote{Electronic address:  don@phys.ualberta.ca},
$^{2,3}$B.~J.~Carr\footnote{Electronic address:  B.J.Carr@qmul.ac.uk},
and $^{4}$Jane H. MacGibbon\footnote{Electronic address:  jmacgibb@unf.edu} }

\affiliation{
$^{1}$Department of Physics, University of Alberta, Edmonton, Alberta T6G 2G7,
Canada\\
$^{2}$Astronomy Unit, Queen Mary, University of London, Mile End Road, London
E1 4NS, UK\\
$^{3}$Research Center for the Early Universe, Graduate School of Science,
University of Tokyo, Tokyo 113-0033, Japan\\
$^{4}$Department of Chemistry and Physics, University of North Florida,
Jacksonville, Florida 32224, USA
}

\date{2008 September 15}


\begin{abstract}

We discuss a variety of bremsstrahlung processes associated with charged
particles emitted by evaporating black holes.  We show that such particles
produce a negligible number of bremsstrahlung photons from their scattering off
each other, though at low frequencies inner bremsstrahlung photons dominate
over the direct Hawking emission of photons.  This analysis and the further
analysis of the accompanying paper invalidate Heckler's claim that sufficiently
hot evaporating black holes form QED photospheres and have similar implications
for putative QCD photospheres.

\end{abstract}

\pacs{04.70.Dy, 12.20.-m, 95.30.Cq, 98.70.Rz
\hfill Alberta-Thy-13-07} 

Phys. Rev. {\bf D 78}, 064044 (2008)

\maketitle

\vspace{-1 mm}

\section{Introduction}

\vspace{-1 mm}

Hawking \cite{H1,H2} has shown that black holes emit thermal radiation, which
for small enough black holes can include massive particles like electrons and
positrons.  Heckler \cite{HE1,HE2} has argued that bremsstrahlung processes
involving such emitted charged particles will produce a large number of photons
for each initial charged particle and lead to a quasithermal photosphere. 
Here we show that the two-body bremsstrahlung (one charged particle scattering
off the electromagnetic field or virtual photons of another charged particle
and emitting one or more free photons) is suppressed by small factors besides
the $\alpha^3$ factor from the minimum number of three photon vertices and is
not enhanced by any large factors such as the gamma factors of the relativistic
charged particles, when the black hole temperature is large compared with the
charged particle rest masses. Therefore, the fraction of the initial charged
particle energy that will go into two-body bremsstrahlung photons is expected
to be of the order of $10^{-8}$ or less.

One of these suppression factors is the ratio of the reduced Compton wavelength
of the emitted particles to their average radial separation. This ratio is
about $5.7\times 10^{-3}$ for relativistic electrons and positrons emitted by a
small hot black hole whose mass is much less than $10^{17}$ grams. Another
factor that may be small for relativistic charged particles (and which goes
inversely with their gamma factor $E/m$) is the causality suppression: The
electromagnetic field of one charged particle emitted by the black hole cannot
scatter another charged particle, leading to the emission of a bremsstrahlung
photon, until there has been time for the electromagnetic field to propagate
causally from the first particle to the second.  This greatly reduces the
classical estimate of the expected momentum transfer between the two particles
from what it would have been if the particles had always existed and had come
in from infinity with the same energies and impact parameter. Thus two-body
bremsstrahlung is suppressed by two small factors (and not enhanced by any
large factors) besides the $\alpha^3$ suppression.  Therefore, the two-body
bremsstrahlung certainly does not seem to be anywhere nearly sufficient to lead
to a photosphere.

There is another bremsstrahlung process that is also small but greater than the
two-body bremsstrahlung process that was incorrectly conjectured to lead to a
photosphere.  This is inner bremsstrahlung, produced by the emitted charged
particles as they change their velocities from initially being effectively at
rest in the frame of the black hole (or a distant observer) to asymptotically
moving radially outward at nearly the speed of light.  This process has only
the single photon vertex of the emitted inner bremsstrahlung photon and so is
suppressed by only one power of $\alpha$.  However, it is enhanced by a
logarithm of the gamma factor of the emitted charged particle.  Therefore, the
inner bremsstrahlung emitted by relativistic electrons and positrons can give
photon luminosities that are tens of percent as large as the direct Hawking
photon emission.  If a black hole is hot enough to emit many massive charged
particle species, then the inner bremsstrahlung emission can have even more
total power than the direct photon emission.

Since the direct photon emission is suppressed at wavelengths long compared
with the size of the hole (giving a very blue spectrum at low frequencies),
whereas the inner bremsstrahlung spectrum is essentially a white spectrum up to
a cutoff near the black hole temperature, the inner bremsstrahlung always
dominates at sufficiently low frequencies.  For example, we shall show below
that a black hole of mass $5.0\times 10^{14}$ grams, whose lifetime equals the
current age of the Universe, has a direct photon total power of about $2.6$
gigawatts, but only about $2.4\times 10^{-29}$ watts is in the visible range
because of the low-frequency suppression. On the other hand, the inner
bremsstrahlung photons from such a hole would give about $4.0$ watts of power in
the visible range, much greater than that of the direct photons in this low-frequency band.

Because the inner bremsstrahlung always greatly dominates over the two-body
bremsstrahlung (which has a relative suppression factor at least as small as
$\alpha^2$ divided by a logarithm of the gamma factor), we may regard inner
bremsstrahlung as the basic bremsstrahlung process and the other bremsstrahlung
processes as merely giving small corrections to it. For example, the two-body
bremsstrahlung may be considered as two-body scattering of the outgoing charged
particles that changes their asymptotic momenta, and hence the resulting
spectrum of inner bremsstrahlung photons, only very slightly. In total,
however, the inner bremsstrahlung always just adds a fraction of order unity to
the direct photon power and does not produce a large number of photons of
energy comparable to that of the charged particles.  Therefore, charged
particle bremsstrahlung cannot lead to a photosphere as Heckler conjectured, as
we shall see by the more detailed arguments below and as is discussed in our
accompanying paper \cite{MCP}.

\vspace{-1 mm}

\section{Two-body bremsstrahlung in terms of impact parameters}

\vspace{-1 mm}

In this paper we shall consider bremsstrahlung by electrons and positrons
emitted by a Schwarzschild black hole of temperature $T_{bh} = (8\pi
M_{bh})^{-1}$ much greater than the electron rest mass $m_e$.  We use Planck
units with $\hbar = c = G = k = 4\pi\epsilon_0 = 1$.  The average emission
number rate of electrons and positrons from such a black hole is \cite{DNP77}
\begin{equation}
 {dN\over dt} \approx 9.516 \times 10^{-4} M_{bh}^{-1},
 \label{1}
\end{equation}
and the average emission power is
\begin{equation}
 {dE\over dt} \approx 1.589 \times 10^{-4} M_{bh}^{-2}.
 \label{2}
\end{equation}
The ratio of the power to the number rate gives the average energy per particle,
\begin{equation}
 \langle E_e \rangle \equiv m_e\gamma_{av}
  = {dE\over dN} \approx 0.167 M_{bh}^{-1} \approx 4.20\, T_{bh}.
\label{3}
\end{equation}
This defines an average gamma factor, $\gamma_{av} \approx 4.20\, T_{bh}/m_e$,
for the emitted electrons and positrons. The size of the black hole, $2M_{bh}$,
is much less than the reduced Compton wavelength of the particles, $1/m_e$. 

We can then write the average emission number rate in terms of $\gamma_{av}$ as
\begin{equation}
 {dN\over dt} = \nu m_e\gamma_{av} \approx 5.70 \times 10^{-3} m_e\gamma_{av},
 \label{4}
\end{equation}
thereby defining the dimensionless numerical constant
\begin{equation}
 \nu \equiv {dN/dt\over dE/dN} = {(dN/dt)^2\over dE/dt}
   \approx 5.70 \times 10^{-3}
 \label{4b}
\end{equation}
that we shall use later.  Our assumption of $T_{bh} \gg m_e$ implies that
$\gamma_{av} \gg 1$, and so almost all of the emitted electrons and positrons are
ultrarelativistic. Because they are moving outward at very nearly the speed of
light, they have an average radial separation in the black hole frame of
$(dN/dt)^{-1}$, which is a factor of $1/\nu \approx 175$ times larger than
their Lorentz-contracted reduced Compton wavelength $1/(m_e\gamma_{av})$.  Thus
the electrons and positrons are, on average, widely separated  once they are
emitted by the black hole.

Since the electrons and positrons have a highly inhomogeneous density
distribution outside the black hole (going essentially as $1/r^2$ for $r \gg
M_{bh}$) and a highly anisotropic momentum distribution (diverging essentially
radially away from the black hole), it is inappropriate to use formulae such as
$\langle n\sigma v\rangle$ for the bremsstrahlung production rate.  Instead, we
shall estimate the expected fraction $F$ of the average charged particle energy
$m_e\gamma$ that goes into bremsstrahlung photons from the electromagnetic
scattering of two ultrarelativistic charged particles emitted by a black hole
with $T_{bh} \gg m_e$.  We will do this by first estimating the corresponding
fraction $f$ from the scattering of an electron or positron (the `scattered
particle') by another individual charged particle (the `scattering particle,'
also emitted by the black hole as part of its Hawking radiation), and then
summing over all such potential scattering particles. 

Note that $f$ and $F$ are expectation values in the quantum mechanical sense
and are {\it not} the expected fraction of the energy of an individual
interacting charged particle that goes into photons when one or more photons
are radiated by that particle.  The emitted photon will typically have an
energy comparable to that of the ultrarelativistic charged particle emitting it
(excluding the infrared divergence in the total number of photons of
infinitesimally small energy) \cite{BH}.  However, for the two-body
bremsstrahlung around the black hole, we find that the probability of the
emission of any such photon is of the order of $\alpha^3 \nu \sim 10^{-8}$.
This is very tiny, so when this probability is multiplied by the fraction of
the charged particle energy carried off by the photon, one gets expected
fractions $f$ and $F$ that are $O(10^{-8})$ or less.  The reason that we shall
focus on the expected energy fractions $f$ and $F$, rather than on the
bremsstrahlung emission probabilities, is that $f$ and $F$, being weighted by
the photon energies, avoid the infrared divergences in the emission
probabilities at infinitesimally small photon energies.

The bremsstrahlung process occurs over a formation length scale $\sim
\gamma/m_e \sim \gamma^2 M_{bh}$ in the center of momentum frame of the
interaction. This is much greater than the black hole radius $2M_{bh}$, so we
shall approximate the black hole as a spatial point (timelike worldline) in
flat spacetime that emits charged particles at the emission rate and power
given above.  For simplicity we shall assume that each charged particle is
emitted with the same energy, corresponding to the gamma factor $\gamma_{av}
\approx 4.20\, T_{bh}/m_e$.  The expected fraction $f$ of the energy of the
charged particle lost as bremsstrahlung photons in Coulomb scattering off
another charged particle will depend on the initial quantum states of the two
particles.  Since we have found that black hole emission leads to particles
widely separated  compared to their Lorentz-contracted reduced Compton
wavelengths, we can use wavepackets in which the particles travel with constant
velocities in straight lines from the point black hole worldline.  Then, with
$\gamma$ having already been assumed, $f$ just depends on the emission angle
$\theta$ between the scattering particle and the scattered particle, and on the
time separation $\delta t$ between the emission of the scattering and the scattered
particle.

We shall approximate $f$ in the $T_{bh} \gg m_e$ black hole case by using the
results of Sec. 4 of Bethe and Heitler \cite{BH} for the bremsstrahlung
radiation probability as a function of impact parameter.  Their calculation is
for a fixed scattering particle (e.g., an atomic nucleus that is assumed not to
recoil), but the results for $f$ should be roughly the same (i.e., to within a
factor of 2 or so) when the scattering and scattered particles have comparable
mass (e.g., when  both are relativistic electrons and/or positrons).  It is
convenient to express their results for the radiation probability as a function
of impact parameter by giving $f$ as a function of the orbital angular momentum
$\ell$ where $\ell$ is the impact parameter multiplied by the linear momentum:
 The linear momentum is roughly $m_e\Gamma$ for $\Gamma \gg 1$, where $\Gamma$
is the gamma factor of the scattered particle in the frame of the scattering
particle.  Then, when the atomic screening is dropped as not relevant for our
case, the Bethe-Heitler Eqs.~(39), (38), and (38A) respectively imply  that the
fraction of the initial energy of the electron emitted as bremsstrahlung
photons is
\begin{eqnarray}
 & f & \sim A \alpha^3 [\ln{(\Gamma/\ell)} + B]
   \quad \mathrm{for} \quad 1 \ll \ell \ll \Gamma;
 \nonumber \\
 & f & \approx (2/\pi) \alpha^3 \Gamma^2/\ell^2
   \quad \mathrm{for} \quad \Gamma \ll \ell \ll \Gamma^2;
 \nonumber \\
 & f & \approx C \alpha^3 \Gamma^4/\ell^3\quad
  \mathrm{for} \quad \Gamma^2 \ll \ell.
 \label{5}
\end{eqnarray}

The numerical constants $A$, $B$, and $C$ are presumably $O(1)$, although they
do not appear to have been explicitly calculated in the literature.  The
normalization of the middle equation has been chosen to make the total cross
section match the dominant contribution of Bethe-Heitler Eq.~(16) in the
ultrarelativistic limit.  One can also deduce the middle equation (though not
the other two) from the results of von Weizs\"{a}cker \cite{vW}.

One can readily calculate that $C = \pi/4$ by the classical formula for the
energy emitted by an accelerating charge $e$ as the time integral of $(2/3)e^2
a^2$, where $a^2$ is the square of the 4-acceleration.  For accelerated motion
in the Coulomb field of another charge $e$ in the limit that the bending angle
is very small, this gives \cite{LL}
\begin{equation}
f \approx (\pi/4) \alpha^3 (\Gamma^2-1/3)(\Gamma^2+1)/\ell^3.
 \label{5b}
\end{equation}

We can then fit all three equations in (2.6) by the following single equation,
after making a simple \textit{ad hoc} choice for $B$ and using
$\ell+1$ instead of $\ell$ in the logarithm to avoid a divergence at $\ell=0$:
\begin{equation}
f(\Gamma,\ell) \sim {2\alpha^3\over\pi c}  
  \ln{\left[1 + c\left({\Gamma\over \ell+1}\right)^2\right]}
     {\pi^2\Gamma^2\over \pi^2\Gamma^2 + 8\ell},
 \label{6}
\end{equation}
where the constant $c$ is to be determined.

Now we can solve for $c$ (and hence $A$) by matching the subleading term
in the energy-averaged total cross section.  From the Bethe-Heitler \cite{BH}
Eq.~(16) for the differential cross section for the emission of a photon of
energy $k$ by an electron of high initial energy $E_0 = \Gamma m_e \gg m_e$,
one can readily calculate that the energy-averaged total cross section is
\begin{equation}
\langle\sigma\rangle = \int\frac{k}{E_0}\frac{d\sigma}{dk}dk \approx
\frac{4\alpha^3}{m_e^2}\left[\ln{(2\Gamma)}-\frac{1}{3}\right],
 \label{6b}
\end{equation}
where all terms with inverse powers of $\Gamma$ have been omitted.  This must
give the same answer as
\begin{equation}
\langle\sigma\rangle
= \frac{\pi}{m_e^2(\Gamma^2-1)}\sum_\ell{(2\ell+1)f(\Gamma,\ell)}
\approx \frac{4\alpha^3}{m_e^2}\left[\ln{\left(\frac{\pi^2}{8}\Gamma\right)}
+\frac{1}{2}-\frac{1}{2}\ln{c}\right].
 \label{6c}
\end{equation}
Equating the right-hand sides of Eqs. (\ref{6b}) and (\ref{6c}) implies that $c
= \pi^4 \mathrm{e}^{5/3}/256$ and $A = 2^9/(\pi^5 \mathrm{e}^{5/3})$ (where
here e is the base of Napierian logarithms and not the magnitude $e$ of the
electron charge), in agreement with our earlier remark that $A$ is $O(1)$. 
Therefore, we can write Eq.~(\ref{6}) with no undetermined parameters as
\begin{equation}
f(\Gamma,\ell) \sim {2^9\alpha^3\over\pi^5 \mathrm{e}^{5/3}}  
  \ln{\left[1 + \frac{\pi^4 \mathrm{e}^{5/3}}{256}
  \left({\Gamma\over \ell+1}\right)^2\right]}
     {\pi^2\Gamma^2\over \pi^2\Gamma^2 + 8\ell}.
 \label{6d}
\end{equation}

Of course, the precise form of this approximation is rather \textit{ad hoc},
such as its inclusion of the addition of 1 to $\ell$ in the logarithm to avoid
a divergence at $\ell=0$, and the rational function used as the final factor,
but we would expect this formula to give the right order of magnitude for all
$\ell$ when $\Gamma\gg 1$.  It would be interesting to do a partial wave
analysis of bremsstrahlung by a Coulomb potential to derive a more precise
approximation in terms of the energy $\Gamma m_e$ of the incoming electron and
its orbital angular momentum $\ell$, especially for $\ell \ll \Gamma$ where the
relative error of Eq. (\ref{6d}) is likely to be largest, though that regime
contributes relatively little in our use of this formula below.

Ignoring factors of the order of unity and logarithms of possibly large numbers
like $\Gamma/(\ell+1)$, the expected fraction of the available energy that goes
into bremsstrahlung photons (essentially the probability that a photon is
emitted with energy comparable to that of the scattered particle) is then $f
\sim \alpha^3$ if the scattered particle gets within its reduced Compton
wavelength $1/m_e$ of the scattering particle, and the fraction drops rather
rapidly with the impact parameter if it is much greater.

\vspace{-1 mm}

\section{Causality suppression of two-body bremsstrahlung}

\vspace{-1 mm}

The Bethe-Heitler bremsstrahlung results are for a charged particle coming in
from infinity and scattering off a stationary Coulomb potential. (The
Bethe-Heitler approach also applies the Born approximation of neglecting
multiple scatterings off the potential, which should be adequate for our
purposes \cite{MCP}.)  However, the bremsstrahlung from the scattering of
particles emitted by a black hole is produced by particles that do not come in
from infinity, but rather are emitted by the black hole.  In our approximation
that the black hole is a point worldline at the spatial origin of flat
spacetime, both the scattering particle and the scattered particle have
worldlines that start at the location of the black hole, rather than coming in
from infinity and passing by each other with some impact parameter.  We are not
certain how to handle this case precisely, but we expect the right order of
magnitude for an upper estimate of the bremsstrahlung simply by replacing the
impact parameter in the Bethe-Heitler results with the minimum distance, $D$,
of the scattered particle from the scattering particle, in the frame of the
scattering particle that replaces the static Coulomb field of the Bethe-Heitler
formula.

Of course, the scattered particle will not detect any influence from the
scattering particle until there is time for a causal signal to travel from the
scattering particle to the scattered particle, after the scattering particle
has been emitted by the black hole.  (Before the causal signal arrives, the
scattered particle would detect the charge of the scattering particle as part
of the charge of the black hole, and we are not considering this here.) 
Therefore, $D$ should be the minimum distance, in the frame of the scattering
particle, to the scattered particle after the scattered particle can receive a
causal signal from the scattering particle.

Let us now calculate this distance $D$.  It will be a function of the gamma
factor $\gamma$ of the particles, the angle $\theta$ between the directions of
the scattering particle and the scattered particle in the black hole frame, and
the time delay $\delta t$ between the emission of the scattering particle and
the scattered particle in the black hole frame (which can be negative or
positive).  We need to join the scattering particle worldline, at spacetime
4-vector position $\mathbf{X}(\tau_1)$ and 4-velocity $\mathbf{u} =
d\mathbf{X}/d\tau_1$, with the scattered particle worldline, at position
$\mathbf{Y}(\tau_2)$ and 4-velocity $\mathbf{v} = d\mathbf{Y}/d\tau_2$, by a
future-directed null line segment $\mathbf{N} = \mathbf{Y}-\mathbf{X}$
(representing the signal carrying the electric field from the scattering
particle to the scattered particle). We then take the dot product of this null
line segment with the scattering particle 4-velocity in order to get the
distance $D = \mathbf{N}\cdot\mathbf{u}$ from the scattering particle to the
scattered particle in the frame of the former.  The result can be most simply
expressed in terms of $\delta t$, $\gamma$, and the relative gamma factor
$\Gamma$ or energy per rest mass of one of the particles in the frame of the
other,
\begin{equation}
\Gamma \equiv \mathbf{u}\cdot\mathbf{v} = \gamma^2 - (\gamma^2-1)\cos{\theta}.
 \label{7}
\end{equation}
Here we have assumed $\gamma_1 \approx \gamma_2 \approx \gamma$, which is true
for the strongly-peaked ultrarelativistic Hawking spectrum.

If the scattered particle is emitted first, so $\delta t < 0$, then the minimum
distance $D$ is obtained by having the null line segment begin on the scattering
particle immediately after it is emitted by the black hole, giving
\begin{equation}
D = \left[(\Gamma-1)\gamma + \Gamma\sqrt{\gamma^2-1}\right](-\delta t).
 \label{8}
\end{equation}
On the other hand, if the scattering particle is emitted first, so $\delta t >
0$, then the minimum distance $D$ is obtained by having the null line segment
end on the scattered particle immediately after it is emitted by the black
hole, giving
\begin{equation}
D = \sqrt{\gamma^2-1}\,\delta t.
 \label{9}
\end{equation}
Therefore, if the minimum distance is to be less than some value
$D_\mathrm{max}$, the  relative difference in the emission times of the
scattering and scattered particles must lie in the range
\begin{equation}
\Delta t = \delta t_\mathrm{max} - \delta t_\mathrm{min}
 = {D_\mathrm{max} \over \sqrt{\gamma^2 - 1}}
     + {D_\mathrm{max}\over (\Gamma-1)\gamma + \Gamma\sqrt{\gamma^2-1}}
 \approx {D_\mathrm{max}\over\gamma}{2\Gamma \over 2\Gamma-1},
\label{10}
\end{equation}
where the approximation applies because $\gamma \gg 1$.  Since $\Gamma \geq 1$
for all values of $\theta$ and $\Gamma \gg 1$ over most of the angular range,
the factor $2\Gamma/(2\Gamma-1)$ lies between 1 and 2 and is usually near 1.
Hence, if we want the minimum distance $D$ between the scattering particle
and the scattered particle in the frame of the former (after there has been
time for a signal to go from the former to the latter and after both have been
emitted from the black hole) to be no greater than some $D_\mathrm{max}$, then
the emission time in the black hole frame of the latter, relative to that of
the former, must occur within a range that is roughly $\Delta t \approx
D_\mathrm{max}/\gamma$.

In particular, for $D$ to be not much greater than $1/m_e$, so that $f$ is not
much smaller than $O(\alpha^3)$, $\Delta t$ must not be much greater than
$1/(m_e\gamma_{av})$. This corresponds to only about $\nu \approx 1/175$ of
the average time between the successive emissions of charged particles. 
Therefore, it would be rare for even one particle to be emitted soon enough to
undergo a scattering that produces a bremsstrahlung photon of significant
energy, even with probability $O(\alpha^3)$, from any putative scattering
particle.

If one uses the approximation of Eq.~(\ref{6d}) for $f$ (though the result
below is dominated by the part where $\Gamma \ll \ell \ll \Gamma^2$, where this
formula is known to be good), assumes that the fraction of the scattered
particle energy lost is roughly the same in the black hole frame (where $F$ is
defined) as in the scattering particle frame (which should be true except for
the small fraction of cases in which both particles are emitted in nearly the
same direction from the black hole), and replaces $\ell$ by $m_e\Gamma D$, then
the total fraction of the scattered particle energy that is emitted into
bremsstrahlung photons becomes
\begin{equation}
 F \approx \int f dN \approx{1\over m_e\gamma_{av}}{dN\over dt}
 \sum_{\ell=0}^{\infty}{1\over\Gamma}f(\Gamma,\ell)
 \approx {2\alpha^3\over m_e\gamma_{av}}{dN\over dt} = 2\alpha^3 \nu
 \approx 4.43\times 10^{-9}.
\label{11}
\end{equation}
This is independent of the black hole mass, provided its temperature is much
greater than the electron mass. (At lower temperatures the charged particle
emission rate, and hence $F$, is exponentially suppressed to even smaller
values by Boltzmann factors in the Hawking distribution \cite{DNP77}.)  As the
black hole becomes hotter than the masses of other species of charged
particles, there will be similar additional contributions from them. However,
provided the number of species is not too large, one will always have $F
\stackrel{<}{\sim} 10^{-7}$. 

Because of the various approximations and {\it ad hoc} assumptions that have
been made to derive our expectation value of the fraction $F$ of the charged
particle energy going into bremsstrahlung from two-particle scattering, we
estimate that it has an uncertainty of at least a factor of 2.  Furthermore,
our approach may give an overestimate by an even greater factor, because the
total momentum transfered by the Coulomb field of the scattering particle to
the scattered particle after they are both emitted by the black hole must be
less than the corresponding momentum transfer in the case of two particles
coming in from infinity, to the same minimum distance, by a factor $\eta$ that
is classically always less, and sometimes much less, than one-half.

To estimate $\eta$ classically, consider for example the case of $\delta t >
0$, so that the scattering particle is emitted first. This case dominates the
estimate for $F$ above. Then when the scattered particle comes out from the
black hole and first receives a signal from the scattering particle, it is, in
the frame of the scattering particle, already at a positive angle $\phi$ beyond
where the position of closest approach to the scattering particle would have
been, if both particles had had constant-velocity worldlines coming in from
infinity.

It is convenient to express this angle $\phi$ in terms of the celerity $p$ (the
spatial distance per proper time, or spatial momentum per rest mass) of a
particle with velocity $v \approx \sqrt{\gamma^2-1}/\gamma$ in the black hole
frame, 
\begin{equation}
 p \equiv \gamma v = \sqrt{\gamma^2-1},
 \label{21}
\end{equation}
the sine of half the
angle between the two particles in the black hole frame,  
\begin{equation}
 s \equiv \sin{\theta\over 2},
 \label{22}
\end{equation}
and the relative velocity (velocity of one particle in the frame of the
other),  
\begin{equation}
 V \equiv {\sqrt{\Gamma^2-1}\over\Gamma} = {2ps\sqrt{p^2s^2+1}\over 2p^2s^2+1}.
 \label{23}
\end{equation}
Then
\begin{equation}
\sin{\phi} = {1\over V}\left[1-{1\over \mathbf{u}\cdot\mathbf{v}}\,
               {\mathbf{N}\cdot\mathbf{v}\over\mathbf{N}\cdot\mathbf{u}}\right]
	   = \sqrt{p^2s^2+s^2\over p^2s^2+1}.
 \label{24}
\end{equation}

If the two particles had come in from infinity with relative velocity $V$ and
with impact parameter $b$, and if one approximates their trajectories as
straight lines with constant velocities, the magnitude of the 4-momentum
transfer classically would be $\Delta P \approx 2e^2/(Vb)$, and $b$ would be
(approximately) the distance of nearest approach.  Similarly, a classical
calculation of the magnitude of the momentum transfer by two particles emitted
from the black hole and thereafter traveling with constant velocity (with $D$
the closest distance of the scattered particle from the scattering particle
once a signal has traveled from the latter to the former), gives the magnitude
of the 4-momentum transfer as
\begin{eqnarray}
\Delta P \approx {2e^2\over VD}\sqrt{{1/2\over 1+\sin{\phi}}-{1\over 4}V^2}
\ \approx\ {e^2\over 2\gamma^2 \delta t}\sqrt{1+\cos{\theta}\over 1-\cos{\theta}},
\label{25}
\end{eqnarray}
where the second approximation of (\ref{25}) applies for $\Gamma = 2p^2s^2+1
\gg 1$ or $1-\cos{\theta} \gg 1/p^2 \approx 1/\gamma^2$.  This implies that the
classical momentum transfer in the black hole case corresponds to that for two
particles coming in from infinity with impact parameter $b = D/\eta$, where the
reduction factor for the magnitude of the 4-momentum transfer is
\begin{equation}
\eta \equiv \frac{VD\Delta P}{2e^2}
\ \approx\ \sqrt{{1/2\over 1+\sin{\phi}}-{1\over 4}V^2}
\ \approx\ \frac{1}{4\gamma}\sqrt{\frac{1+\cos{\theta}}{1-\cos{\theta}}}.
\end{equation}
Replacing $\ell$  by $m_e\Gamma b = m_e\Gamma D/\eta$, rather than by
$m_e\Gamma D$, we then get the following smaller estimate for the total
fraction of the scattered particle energy that is emitted into bremsstrahlung
photons:
\begin{equation}
F \approx {\pi \alpha^3 \nu\over 4\gamma_{av}}.
 \label{27}
\end{equation}

Thus this classical estimate of $F$ is suppressed by a factor of
$\pi/(8\gamma_{av})$, i.e., by the inverse of the charged particle average
gamma factor. This suggests that the average fractional energy loss to
bremsstrahlung may be significantly lower than the already small estimate of
$4\times 10^{-9}$ of Eq. (\ref{11}).  However, if we include the finite size of
the black hole and nonradial motions of the emitted particles near the hole,
the fractional energy loss is probably decreased by less than the gamma factor
of (\ref{27}), though it may still be significantly lower than the conservative
upper bound given by Eq. (\ref{11}).

Since our estimate for $F$ given by Eq. (\ref{11}), which should be taken as a
rough upper limit on the average fraction of energy lost to two-particle
bremsstrahlung by a charged particle emitted from a $T_{bh} \gg m_e$ black
hole, is more than 8 orders of magnitude smaller than unity, it strongly
indicates that the two-body bremsstrahlung radiated by the scattering of the
charged particles emitted by a black hole is completely negligible and
insufficient to form a photosphere by many orders of magnitude.  This strong
suppression comes primarily from our inclusion of the causality constraint
which was omitted in the Heckler photosphere model.  Applications to QCD are
discussed in \cite{MCP}.

The causality suppression that we have calculated here may also be viewed as a
partial justification of the approximation used in \cite{ADO} of only including
scattering of partons that approach each other.  That paper concluded that TeV
higher-dimensional black holes that might be created by the CERN LHC will not form
chromospheres.

\vspace{-1 mm}

\section{Inner bremsstrahlung emission}

\vspace{-1 mm}

There are several other radiative processes that are also generally smaller than
the direct photon emission but comparable to or larger than the two-particle
bremsstrahlung analyzed above.  Most important is inner bremsstrahlung
\cite{KU,Bloch,Jac} from the charged particles that are emitted from the black
hole. To an observer at infinity, each charged emitted particle appears to have
its velocity change from rest at the position of the hole to an
ultrarelativistic outward asymptotic velocity.  The change in the
electromagnetic field of the emitted particle from the Coulomb field
corresponding to the initial velocity to the Coulomb field corresponding to the
final velocity should appear to a distant observer as inner bremsstrahlung
radiation. The analogous calculation of electromagnetic radiation from an
individual relativistic charged particle falling into a neutral black hole was
performed in Ref.~\cite{CLY}.

In the black hole frame, it is most probable that the inner bremsstrahlung
photon will be emitted in very nearly the same direction as that of the charged
particle that emits it.  If the photon carries a fraction $f$ of the energy
$m_e\gamma$ of the charged particle in the black hole frame and hence has
wavelength $(f m_e\gamma)^{-1}$, it will become separated from the charged
particle by one wavelength after traveling outward by a distance $\sim (f
m_e\gamma)^{-1}/(1-v_e) \sim \gamma/(f m_e)$ from the black hole.  This exceeds
the size of the hole, $\sim M_{bh} \sim 1/T_{bh} \sim 1/(m_e\gamma)$, by a
factor of roughly $\gamma^2/f$.  This distance, roughly  $\gamma^2 M_{bh}/f$,
is the effective formation length for the bremsstrahlung photon \cite{Klein}. 
Since it is so much larger than the size of the black hole (for $\gamma \gg
1$), we may simply apply the standard flat spacetime analysis
\cite{KU,Bloch,Jac}.

The inner bremsstrahlung process involves just one photon vertex, so the
fraction of the particle's energy expected to be radiated (i.e., the probability
that a charged particle of a given energy will emit a photon of comparable
energy) is $O(\alpha)$.  This is small, but not nearly so small as the
$O(\alpha^3)$ bremsstrahlung analyzed above arising from the interaction of two
charged particles which involves three photon vertices. The total power in the
inner bremsstrahlung photons will be less than the power in the direct photon
emission by a factor $O(\alpha)$. Since the direct photon emission has a power
per frequency interval that goes as the fourth power of the frequency at low
frequencies \cite{DNP76}, whereas the inner bremsstrahlung photons have a white
spectrum at frequencies well below the black hole temperature, the white inner
bremsstrahlung photons will dominate at photon energies below $\sim \alpha^{1/4}
T_{bh}$.

More precisely, the number flux of inner bremsstrahlung photons of energy
$\omega$ radiated by particles of mass $m$ and charge $\pm e$ that were emitted
from the black hole with a spectrum $d^{2}N_\epsilon/dtdE$ is \cite{KU}
\begin{eqnarray}
{d^{2}N_{b\gamma}\over dt d\omega}
 &=& {2\alpha\over\pi\omega} \sum_\epsilon 
 \int_m^\infty dE {d^{2}N_\epsilon\over dt dE}
  \left[{E^2+2\epsilon m(E-\omega)+(E-\omega)^2
   \over 2(E+\epsilon m)\sqrt{E^2-m^2}}
    \ln{E-\omega+\sqrt{(E-\omega)^2-m^2} \over m}
  -{\sqrt{(E-\omega)^2-m^2} \over \sqrt{E^2-m^2}}  \right]
 \nonumber \\
 &\approx& {2\alpha\over\pi\omega}
 \left[\ln{(2\gamma_{av})}-1\right]{dN\over dt},
 \label{12}
\end{eqnarray}
where $\epsilon = 2(j-\ell) = \pm 1$ for total angular momentum $j$ and orbital
angular momentum $\ell$, and where the approximation of the last line of Eq.
(\ref{12}) applies for $\gamma_{av} \approx E/m \gg 1 + \omega/m$.  Multiplying
Eq. (\ref{12}) by $\omega$ gives the corresponding spectrum for the power of
the inner bremsstrahlung photons. The power spectrum from a charged particle of
initial energy $E$ is very nearly flat until the photon energy $\omega$
approaches its maximum value of $E-m$, where the spectrum rapidly but smoothly
drops to zero. The total power radiated in inner bremsstrahlung photons by
charged particles emitted from the black hole with power $dE/dt$ is then
\begin{eqnarray} {dE_{b\gamma}\over dt} &\approx& {2\alpha\over\pi}
\left[\ln{(2\gamma_{av})}-1\right]{dE\over dt}.
\label{13}
\end{eqnarray}
The mean gamma factor $\gamma_{av} \approx 4.20 T_{bh}/m$ is given by Eq.
(\ref{3}) for ultrarelativistic charged spin-half particles emitted by a  black
hole of temperature $T_{bh} \gg m$ (or $M_{bh} \ll 10^{17}(m_e/m)$ g).  

We may then use Eqs. (\ref{1}) and (\ref{2})  to compare the inner
bremsstrahlung photon power $dE_{b\gamma}/dt$ with the direct photon power
given in \cite{DNP76} and \cite{DNP76b}, which is
\begin{equation}
{dE_{d\gamma}\over dt} \approx 0.3364\times 10^{-4} M_{bh}^{-2}.
 \label{14}
\end{equation}
For example, for a black hole with $T_{bh} = 50$ GeV, the gamma
factors are $411\,000$, 1986, and 118 respectively for electrons, muons, and
taus, and so the sum of the three logarithmic square-bracket factors of Eq.
(\ref{13}) is 24.37.  For the ratio of the inner bremsstrahlung photon power to
the direct photon power, one must multiply this logarithmic factor by
\begin{equation}
{2\alpha \over\pi} {dE/dt\over dE_{d\gamma}/dt}
\approx (0.004646)(4.724) \approx 0.02195,
 \label{15}
\end{equation}
where $dE/dt \approx 1.589 \times 10^{-4} M_{bh}^{-2}$  from Eq. (\ref{2})
represents the power in each species of ultrarelativistic spin-half charged
particle and antiparticle, and $dE_{d\gamma}/dt \approx 0.3364 \times 10^{-4}
M_{bh}^{-2}$ from Eq. (\ref{14}) represents the power in photons directly
emitted by the Hawking process.  For a black hole with $T_{bh} = 50$ GeV,
one then finds that the inner bremsstrahlung photons radiated by the electrons,
muons, taus, and their antiparticles give 53\% as much power as the directly
emitted photons.

Thus for such a hot black hole, the sum of the logarithm factors (24.37),
multiplied by the factor 4.734 for the greater power in spin-half particles and
antiparticles than in direct photons (a factor of 2.362 from  the lower
centrifugal barrier for a black hole to emit particles of lower spin
\cite{DNP76,DNP76b,DNP77}, multiplied by a factor of 2 for the inclusion of
distinct antiparticles for the spin-half particles), nearly compensates for the
small factor of $(2/\pi)\alpha \approx 0.004646$ from the single photon vertex
of this inner bremsstrahlung process. When one includes the emission of other
charged particles, such as pions and/or quarks, the total power in inner
bremsstrahlung photons could possibly exceed the total power in photons
directly emitted by the Hawking process for a sufficiently hot black hole.

If one considers a cooler black hole, say with temperature $T_{bh} \approx 21$
MeV and mass $M_{bh} \approx 5.0\times 10^{14}$ g, whose lifetime equals the
present age of the Universe \cite{DNP76,MCP,M2}, then only electrons and
positrons are ultrarelativistic, with $M_{bh} m_e \approx 0.0010$.  Muons and
antimuons are partially relativistic, with $M_{bh} m_\mu \approx 0.20$, and
taus have $M_{bh} m_\tau \approx 3.34$ or $m_\tau /T_{bh} = 8\pi M_{bh} m_\tau
\approx 84$ and so are hardly emitted at all.  In this case the electrons and
positrons have $\gamma_{av} \approx 173.7$, while the muons and antimuons have
$\gamma_{av} \approx 1.305$.  The power in the ultrarelativistic electrons and
positrons is given by Eq. (\ref{2}). Reference~\cite{DNP77} similarly gives the
power in muons and antimuons when $M_{bh} m_\mu \approx 0.20$ as
\begin{equation}
 {dE\over dt} \approx 0.491 \times 10^{-4} M_{bh}^{-2}.
 \label{16}
\end{equation}
The power in muons and antimuons is only about 30\% as large as for electrons
and positrons because of the Boltzmann suppression in the Hawking distribution
due to the greater mass of the muon.  For $M_{bh} \approx 5.0\times 10^{14}$ g,
the ratio of inner bremsstrahlung to direct photon power is then
\begin{equation}
{dE_{b\gamma}/dt \over dE_{d\gamma}/dt} \approx 0.108,
 \label{17}
\end{equation}
of which nearly 99\% comes from the inner bremsstrahlung photons generated by
the electrons and positrons, and about 1.3\% comes from the muons and
antimuons.

At low frequencies, the spectrum for the power per frequency interval from
direct photons in the Hawking radiation is very blue, going as the fourth power
of the frequency \cite{DNP76}, whereas the inner bremsstrahlung spectrum is
independent of the frequency (white), up to the cutoff at the energy of the
emitting charged particles.  Therefore, at sufficiently low frequency, the
inner bremsstrahlung photons dominate over the direct photons.  Using the
results of Refs.~\cite{DNP76} and \cite{DNPthesis}, one can see that for
$M_{bh} \approx 5.0\times 10^{14}$ g, the inner bremsstrahlung photon spectrum
dominates for $M_{bh}\omega < 0.107$, that is for photon energies $\omega < 57$
MeV, whereas the peak of the direct photon spectrum occurs at $M_{bh}\omega
\approx 0.24$ and $\omega \approx 130$ MeV.

Equation (\ref{13}) implies that the power spectrum of inner bremsstrahlung photons
from a black hole of mass $M_{bh} \approx 5.0\times 10^{14}$ g is 
\begin{equation}
{d^{2}E_{b\gamma}\over dt d\omega} \approx 1.73\times 10^{19} \, {\mathrm s}^{-1}.
 \label{18}
\end{equation}
The bremsstrahlung photons are cut off above an energy which is roughly the
average energy of the electrons and positrons, about $4.20 T_{bh} \approx 90$
MeV from Eq. (\ref{3}). If one integrates over the frequencies of photons in
the visual range, say between 400 and 750 nm, one finds that the power in
visible photons (almost entirely inner bremsstrahlung radiation) is about $4.0$ W. The total power in all frequencies is about 2.56 gigawatts (about $6.4 \times 10^{8}$
times greater than that in the visible range) and peaks at around 130
MeV, where the directly emitted Hawking photons dominate.  However, these
direct photons have a spectrum of the form \cite{DNP76}
\begin{equation}
{d^{2}E_{d\gamma}\over dt d\omega} = {8\over 3\pi^2} M^3 \omega^4
\label{19}
\end{equation}
at low $\omega$, giving a power in the visible range of only about $2.4\times
10^{-29}$ W. This is about $6.0\times 10^{-30}$ times that of the inner
bremsstrahlung photons in the visible range and is less than $10^{-38}$ of the
total photon power.

Since a star of visual magnitude $m_V=6$, which is barely visible to the naked
human eye, gives a visible photon energy flux of about $10^{-8}$ lux
\cite{Allen}, where a lux is a lumen per square meter and a lumen at the
visible wavelength 555 nm is 0.001\,47 W \cite{JM}, a barely visible
light source emits a flux of visible photons of about $1.5\times 10^{-11} \,
\mathrm{W}\mathrm{m}^{-2}$.  Therefore, for a $5\times 10^{14}$ g black hole,
generating 4 W of power in the visible spectrum by inner bremsstrahlung, to
be just visible to the human eye, it must be within a distance of about 150
km.  However, at that distance, there will be a flux of about $0.01 \,
\mathrm{W}\mathrm{m}^{-2}$ of high-energy gamma rays.  This would lead to the
recommended maximum yearly dosage in a human, of the order of 1 rem \cite{PVM},
which deposits an energy of $0.01 \, \mathrm{J}\mathrm{kg}^{-1}$ (see p. 120 of Ref.
\cite{JM}) in a time of the order of 4 minutes.  In roughly half a day, one
would receive about 200 rem, which ``will cause vomiting in 50\% of those
exposed after about 3 hours,'' and in roughly a full day, one would receive
about 450 rem, which is ``the radiation dose that gives a 50\% probability of
death \ldots for healthy people without medical treatment'' \cite{Cohen}. 
Therefore, one would not want to stay unprotected very long close enough to a
$5\times 10^{14}$ g black hole to be able to see it without a telescope.

The proposed 100-meter Overwhelmingly Large Telescope (OWL), whose concept is
now being studied by the European Southern Observatory, ``will be able to reach
magnitude 38 in 10 hours exposure time.  This is a factor five thousand billion
\ldots fainter than the faintest star visible to the naked eye'' \cite{OWL}. 
In principle OWL should see the inner bremsstrahlung visual photons from a
$5\times 10^{14}$ g black hole at a distance of about $4\times 10^{8}$ km or
so, or about 3 AU, roughly the distance from Earth to the Asteroid Belt.  At
this distance the gamma ray flux would be less than $2\times
10^{-15} \, \mathrm{W}\mathrm{m}^{-2}$, which would be quite safe for humans even
without shielding.

\vspace{-1 mm}

\section{Other bremsstrahlung processes}

\vspace{-1 mm}

Besides the inner bremsstrahlung that comes from the change as viewed from
infinity in the velocity of a charged particle as it is emitted out of a black
hole, another effect is the bremsstrahlung emitted when a charged particle
coming out of the black hole scatters off the electric field of the hole
itself.  As the black hole emits electrons and positrons stochastically, it
will have an rms charge $O(1)$ in Planck units. (In Planck units with
$4\pi\epsilon_0=1$, the positron charge is $e = \alpha^{1/2}$.)  For example,
Ref.~\cite{DNP77} showed that for a black hole of $T_{bh} \ll m_e$, the rms
charge is approximately $1/\sqrt{8\pi} \approx 0.1995 \approx 2.335 e$, whereas
for $T_{bh} \gg m_e$, as is applicable for our calculations, the rms charge is 
$\approx 0.5247 \approx 6.143 e$.  The average square of the photon vertex with
the black hole is then of order unity.  However, to calculate the fraction of
the power in these bremsstrahlung photons, there will be one photon vertex on
the scattering particle involving the Coulomb field of the black hole and one
vertex involving the emitted bremsstrahlung photon. Thus the fraction of the
power will be $O(\alpha^2)$.  This is one power of $\alpha$ smaller than the
inner bremsstrahlung but is still one power of $\alpha$ larger than the
two-particle bremsstrahlung analyzed in Secs. II and III.

Another bremsstrahlung process is the scattering by a charged particle of the
inner bremsstrahlung photon emitted by another charged particle. This process
gives outgoing photons of expected power fraction $O(\alpha^3)$ and so would be
comparable to the two-particle bremsstrahlung analyzed above, but two powers of
$\alpha$ smaller than the inner bremsstrahlung itself.  This process could be
described classically by noting that the retarded field of the scattering
particle does not have a purely Coulomb form (corresponding to constant
velocity of the scattering particle) at the retarded time when it is just
coming out from the black hole. As it is being emitted, the charged particle is
effectively accelerated from being at rest at the black hole position
(approximating the black hole as a point in flat spacetime) to having its
asymptotic post-emission velocity. Since this scattered bremsstrahlung (the
photons emitted by the inner bremsstrahlung and then scattered by another
charged particle) is $O(\alpha^2)$ smaller than the inner bremsstrahlung
directly emitted by the scattering particle, it is already known to be small.
Thus we can ignore it and take the bremsstrahlung from the scattering particle
to be that found above by assuming the particle to have achieved its asymptotic
velocity and hence to have a retarded field which is purely Coulombic, with
electric field $e/d^2$ where the distance $d$ is measured in the frame of the
scattering particle ($D$ being the minimum value of $d$).

There are also other radiative effects, discussed in Ref.~\cite{DNP77}, of
smaller order, such as the vacuum polarization by the black hole with its
fluctuating charge, and self-energy corrections to the propagation of the
charged particles.  The latter effect comes from the fact that the charged
particles will be surrounded by clouds of virtual photons and so will not
propagate in the black hole spacetime exactly the same way as the point Dirac
particles numerically analyzed in Ref.~\cite{DNP77}.

\vspace{-2 mm}

\section{Discussion}

\vspace{-2 mm}

All of these effects analyzed in this paper are generally small in comparison
with the direct emission of particles and photons from the black hole (except
in comparison with photons well below their peak frequency), and none of these
effects will lead to photospheres.  In particular we have confirmed that the
two-particle bremsstrahlung is far too weak to lead to a QED photosphere.

An intuitive way of seeing our result, that the fraction of energy going into
two-body bremsstrahlung photons is not enhanced above the $\alpha^3$ value that
one would get from simply counting photon vertices, even for large gamma
factors, is as follows:  If one regards each electron as being a
three-dimensional blob of radius $1/m_e$ (its Compton radius) in its own rest
frame, then as the electrons are emitted by a black hole with gamma factors
$\gamma \sim T_{bh}/m_e \gg 1$ and their Coulomb fields propagate at finite
speed, the electrons are Lorentz-contracted in the radial direction to
thickness $\sim 1/(m_e\gamma)$ and their shapes distorted to resemble eggshells
in the black hole frame.  This thickness is less than the average radial
separation between the electrons being emitted by the black hole by a factor of
about $1/\nu \approx 175$.  (This factor is a number of order unity in the
sense that it contains no powers of any physical quantities like $\alpha$ or
$m_e$ but just comes from numerical solutions \cite{DNP77} of the Dirac
equation around a black hole.)  Roughly speaking, bremsstrahlung occurs with
probability $\sim \alpha^3$ if two electron blobs overlap, but since the
overlap has low probability $\sim \nu \approx 5.70 \times 10^{-3}$ here, the
probability is even lower than $\alpha^3$.  In particular, there is no
enhancement when $\gamma$ gets large, since the factor of $1/\gamma$ in the
Lorentz contraction precisely cancels the factor of $\gamma$ in the emission
rate.

Another way of expressing the smallness of the two-body bremsstrahlung is the
following:  Because the bremsstrahlung formation length, given in Sec. IV as
$\sim \gamma^2 M_{bh}/f$, is so much larger than the size of the black hole,
what is most relevant for the spectrum of the bremsstrahlung photons is the
change in the momentum of each charged particle over this distance.  Since
each charged particle effectively starts at rest at the black hole from the
perspective of a viewer at infinity, the change in its spatial momentum over
this distance is very nearly its asymptotic momentum.  Therefore, the
bremsstrahlung spectrum (including the inner bremsstrahlung and the various
corrections to it from the scattering from the fluctuating charge of the black
hole and from other charged particles that are emitted) can be calculated
purely from the asymptotic momentum distribution of the charged particles
emitted by the black hole. A very good approximation for this momentum
distribution is given simply by using the Hawking emission formula with
numerical solutions for the particle wavefunctions propagating in the field of
the black hole, such as was done for charged spin-half particles in
Ref.~\cite{DNP77}.  In this approach, the bremsstrahlung can be regarded purely
as inner bremsstrahlung with fractional energies of $O(\alpha)$.  The Coulomb
scattering of the charged scattered particles from other charged scattering
particles in the outgoing flux then gives $O(\alpha^2)$ corrections to the
asymptotic momentum distribution and hence to the bremsstrahlung spectrum,
leading to an absolute effect of $O(\alpha^3)$ for the fraction of energy going
into bremsstrahlung photons.  From this viewpoint, the scattering between
different charged particles emitted by the black hole gives only a tiny
$O(\alpha^2)$ correction to the inner bremsstrahlung and so is effectively 
negligible.

Our results here confirm what was written in Ref.~\cite{DNP77}, ``Since the
average time between the emission of successive leptons will turn out to be
greater than $10^3 M$, it should be a very good approximation to ignore the
interactions between different leptons emitted.''  In particular, the
interactions between charged particles emitted by a black hole are almost
completely negligible and cannot form a QED photosphere as Heckler \cite{HE1,
HE2} has claimed.  Similar implications for QCD chromospheres are discussed in
\cite{MCP,ADO}.

\vspace{-2 mm}

\begin{acknowledgments}

\vspace{-2 mm}

We are grateful to Andrzej Czarnecki and Manuel Drees for helpful discussions. 
The research of DNP was supported in part by the Natural Sciences and
Engineering Research Council of Canada. BJC thanks the Research Center for the
Early Universe at the University of Tokyo for hospitality received during this
work. JHM thanks the University of Cambridge for hospitality received during
this work.

\end{acknowledgments}

\end{document}